# Synthesis and SERS application of SiO$_2$@Au nanoparticles

*A. Saini, T. Maurer\*, I. Izquierdo Lorenzo, A. Ribeiro Santos, J. Béal, J. Goffard, D. Gérard, A. Vial and J. Plain*

*Laboratoire de Nanotechnologie et d'Instrumentation Optique, ICD CNRS UMR n°6281, Université de Technologie de Troyes, CS 42060, 10004 Troyes, France*

In this letter, we report a chemical route for synthesizing SiO$_2$@Au core-shell nanoparticles. The process includes four steps: i) preparation of the silica cores, ii) grafting gold nanoparticles over SiO$_2$ cores, iii) priming of the silica-coated gold nanoparticles with 2 and 10 nm gold colloids and finally iv) formation of complete shell. The optical extinction spectra were experimentally measured and compared to numerical calculations in order to confirm the dimensions deduced from SEM images. Finally, the potential of such core-shell nanoparticles for biosensing was probed by means of Surface Enhanced Raman Scattering measurements and revealed higher sensitivities with much lower gold quantity of such core-shell nanoparticles compared to Au nanoparticles exhibiting similar diameters.

SiO$_2$ core-Au shell nanoparticles have attracted much attention for the past ten years due to potential applications in various fields such as cancer imaging and treatment(Loo et al., 2004), Surface-Enhanced Raman Spectroscopy (SERS) detection of molecules(Wang et al., 2006, Maurer et al., 2013), catalytic degradation of environmental pollutants(Ma et al., 2009) or even fabrication of SPASER (Surface Plasmon Amplification by Stimulated Emission of radiation) devices when the silica is dye-doped(Stockman, 2008). However, the growth of a complete gold shell around silica nanoparticles is very hard to achieve. Recently, most reports have been limited to the grafting of many Au small colloids onto silica cores(Hiramatsu and Osterloh, 2003, Osterloh et al., 2004, Shi and Asefa, 2007, Zhang et al., 2007). In this case, the more Au colloidal particles surround silica cores, the more red-shifted are their Localized Surface Plasmon Resonances (LSPR), which are typically between 500 nm and 550 nm. The optical signature of a complete gold shell depends in fact on the SiO$_2$ core diameter. When the SiO$_2$ cores are large enough (typically larger than 150 nm), the completion of the gold shell is evidenced by the apparition of two LSPR modes in the visible and near-infrared range(Oldenburg et al., 1998), which both blue-shift when the gold shell thickness increases. However, when the SiO$_2$ cores exhibit lower diameters, only one LSPR peak can be observed in the extinction spectra, which also blue shifts with larger shell thickness. Surprisingly, it seems that there has been no successful report of complete shell growth since the Halas and coworkers study fifteen years ago(Oldenburg et al., 1998). The basics of the synthesis process was indicated but not detailed: i) synthesis of the SiO$_2$ cores via the Stöber method(Stöber et al., 1968), ii) grafting of previously prepared very small gold colloids(Loo et al., 2004) via organosilane molecules and iii) growth of additional gold onto the colloids via chemical reduction. Therefore, the complete gold shell synthesis around silica cores is far from being easy to achieve.

In this paper, we report a detailed process that allows growing a complete gold shell around silica cores. This process presents 4 steps: i) preparation of the silica cores, ii) grafting gold nanoparticles onto SiO$_2$ cores, iii) priming of the silica-coated gold nanoparticles with 2 and 10 nm gold colloids and finally iv) formation of complete shell. We present Scanning Electron Microscopy (SEM) images and optical extinction measurements compared to Mie calculations in order to prove the efficiency of this method. We also propose the use of such core-shell nanoparticles for SERS applications since they provide higher sensitivity with much lower gold materials than gold colloids.

**Experimental section**

**Materials**

The following materials were used without further purification for the procedure. Ammonium hydroxide (Fisher, 25%), tetraethyl orthosilicate (TEOS) (Sigma Aldrich,98%), (3-aminopropyl)trimethoxysilane (Sigma Aldrich,98%), hydrogen tetrachloroaurate(III) hydrate (HAuCl4)(Sigma Aldrich,99.99%), Sodium hydroxide (NaOH) (Sigma Aldrich), formaldehyde (Sigma Aldrich), Potassium Carbonate (K2CO3) (Sigma Aldrich). Ultrapure deionized water (continental Water Systems) was used for all solution preparations and experiments. All glasswares were treated with sulfuric acid digestant and then cleaned with acetone, ethanol and deionized water.

---

[*] To whom correspondence should be send to: Thomas Maurer (thomas.maurer@utt.fr)

**i) Preparation of the silica cores**

Ammonium hydroxide (25%, 2.5 mL) was added to absolute ethanol (50 mL). While the above mixture was stirred vigorously, TEOS (98%, 1.5 mL) was added dropwise. After 6 hours, the opaque white $SiO_2$ nanoparticles were obtained. The solidified silica nanoparticles were collected by centrifugation at 3000 rpm for 10 min and washed three times with ethanol. Between the washes, the nanoparticles were dispersed by vortex and sonication.

**ii) Grafting gold nanoparticles onto $SiO_2$ cores**

Firstly, the $SiO_2$ nanoparticles had to be functionalized. A 0.5 mL solution of 60 mM (3-aminopropyl)trimethoxysilane (APTMS) in EtOH:deionized water (3:1 volume ratio) was added dropwise into the $SiO_2$ solution under sonication for 3 hours. The functionalization was monitored visually by observing the separation of the solution into two layers when left unstirred: the APTMS-coated silica nanoparticles precipitated on the bottom, leaving a clear ethanolic solution at the top. To enhance the covalent bonding between the (3-aminopropyl)trimethoxysilane groups and the silica shell surface, the solution was gently refluxed for one additional hour before being purified by centrifugation at 2000 rpm for one hour and redispersed in 5 mL of water. This functionalization step was repeated twice to further improve the silanization reaction. Then, the grafting of gold colloids could start. A solution of $HAuCl_4$ (10 mL, 50 mM) together with a solution of NaOH (5 mL, 0.1 M) were added to the APTMS-grafted silica spheres to adjust the pH value to 7-7.5, and the solution was left under stirring for 15 minutes to incorporate the gold seeds into the spheres. After that, we could see the white $SiO_2$ particles showing a yellow color. The solution was centrifuged at 2500 rpm and washed 3 times with deionized water. After that, it was re-dispersed ultrasonically in deionized water and diluted to 100 mL with deionized water. It was heated to reflux after which a solution of formaldehyde (2.5 mL, 38.8 mM) was rapidly added maintaining the refluxing conditions to reduce the grafted gold salt to nanoparticle range. The color changed to red and the solution was then centrifuged at 2500 rpm 3 times in purified water.

**iii) Priming of the silica-coated gold nanoparticles with 2 and 12 nm gold colloids**

A 500 μL solution of 1 mM APTMS was added to a 5 mL aliquot of gold-coated silica nanoparticles silica-coated gold nanoparticles under vigorous stirring. The mixture was allowed to react for 2 h, at room temperature under argon, followed by the addition of a 5 mL colloidal suspension of 2 nm gold nanoparticles (prepared by Duff's method(Duff et al., 1993)). The resulting solution was kept under gentle stirring overnight. The mixture was then centrifuged at 2500 rpm for 1 h. The supernatant was decanted, leaving a dark colored pellet, which was re-dispersed and sonicated in 2.5 mL of purified water. This step was then followed by priming the gold coated silica nanoparticles nanoparticles with 12 nm gold colloids (synthesized via Turkevich's method(Turkevich, 1985b, Turkevich, 1985a)). The priming method was the same as that described above for the 2 nm gold colloids.

**iv) Formation of the complete shell**

After the priming of the silica shells was completed, the solution was left undisturbed for 48 hours to ensure efficient priming by Ostwald ripening(Wynblatt, 1976, Wynblatt and Gjostein, 1976). For obtaining a complete gold shell around silica nanoparticles, potassium-gold solution (K-gold) is required which was prepared by mixing 100 mL aqueous $K_2CO_3$ solutions (280 mg solid/L) with a stock solution of $HAuCl_4$ (1.5 mL, 25 mM) by continuous stirring and aging in the dark for 12 h. To initiate the growth of a complete gold shell, the precursor silica particles covered with the small gold clusters were added to the aged $HAuCl_4/K_2CO_3$ solution. The ratio between precursor particles and gold salt solution depended on the intended thickness of the gold shells. It was calculated assuming that all added $HAuCl_4$ would be reduced to yield gold shells around the silica particles. Gold nanoshells were pre-pared by varying the K-gold:gold-seeded silica volume ratio from 200 to 500:1 using an aqueous solution of $NaBH_4$ (5.3 mM) as the reducing agent.

**Results**

SEM images were performed using a FEG eLine Raith (10 kV) with an aperture opening of 30 μm. $SiO_2$ nanoparticles with two different sizes were investigated. Thanks to a statistical analysis of the images, the average diameter of the smaller $SiO_2$ cores was assessed to be 88 nm with a standard deviation of 11 nm (see **Fig. 1A**) while the larger ones are about twice larger since they exhibit a mean diameter of 170 nm with a standard deviation of 18 nm. The analysis of SEM images after the gold nanoshell growth clearly shows an increase of the average diameter to 128 nm with a standard deviation of about 34 nm for the smaller $SiO_2$ nanoparticles (see **Fig. 1B**), which gives evidence that a complete shell was grown. Moreover, it can also be observed that gold colloids were formed during the synthesis due to the reduction of gold salts. These gold colloids can be removed after centrifugation. The other proof for the complete gold shell growth comes from SEM and TEM images of isolated nanoparticles (**Figs 1C** and **1D**, respectively), which show facetted shapes of the core-shell nanoparti-cles. As for the larger SiO$_2$ nanoparticles, the diameter increase due to the shell growth was limited to about 10 nm and the shell surface looks rather rough (see **Fig. 1F**).

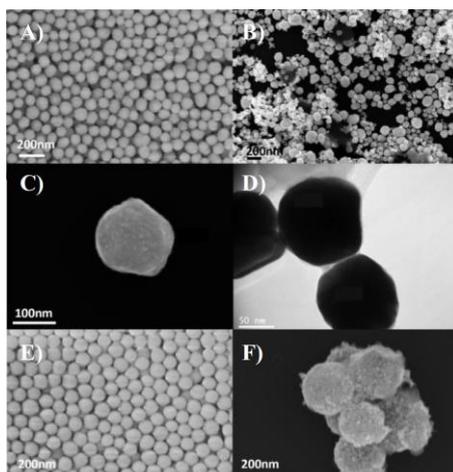

**Fig. 1** A) SEM image of SiO$_2$ nanoparticles before gold nanoshell growth. The average diameter is 88 nm with a standard deviation of 11 nm. B) SEM images of core-shell nanoparticles once the process is completed. Small gold colloids, observed on the image, can be removed after centrifugation. The average diameter of the core-shell nanoparticles is 128 nm with a standard deviation of 34 nm. C and D) SEM and TEM images of an isolated core-shell nanoparticle. The gold shell growth leads to facetted shapes. E) SEM image of larger SiO$_2$ nanoparticles before gold nanoshell growth. The average diameter is 170 nm with a standard deviation of 18 nm. F) SEM images of isolated core-shell nanoparticles after complete shell growth from the larger SiO$_2$ nanoparticles shown in Figure 1E.

In order to confirm the growth of a complete gold shell, and not only gold colloid grafting, visible-IR spectra were per-formed on the core-shell nanoparticles deposited by evaporation at room temperature, on a glass substrate (see **Fig. 2**).

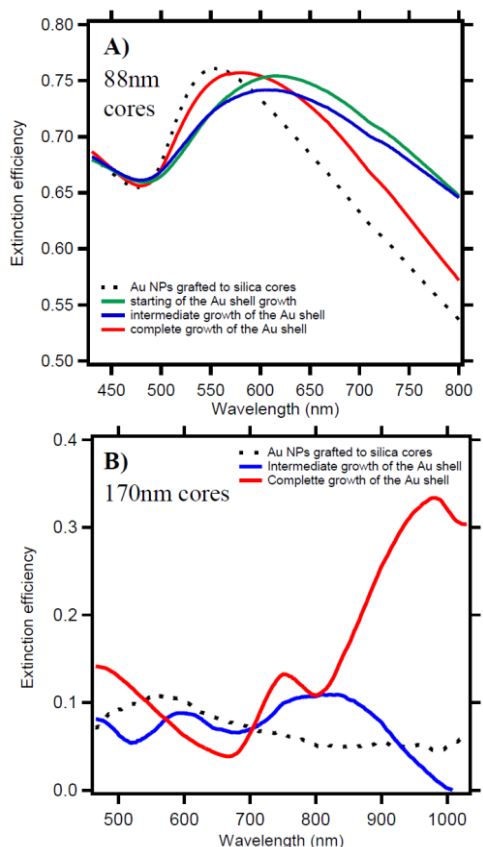

**Fig. 2** Visible-IR spectrum of core-shell nanoparticles deposited on a glass substrate for the smaller (A) and larger (B) SiO$_2$ cores. After priming of the silica cores with 2 and 10 nm Au NPs, a LSPR peak is observed

around 550 nm (dashed dark curves). A) For the smaller SiO$_2$ cores, the growth of the gold shell leads to a peak around 624 nm (green curve) which then blue-shifts while the shell is growing (blue and red curve). The extinction spectrum (red curve) corresponding to the SEM images (see **Fig. 1**) provides a LSPR wavelength around 584 nm. B) For the larger SiO$_2$ cores, the growth of the gold shell leads to two peaks around 977 nm and 747 nm (red curve). Once the growth is completed, these two peaks are broadened and blue-shifted to 795 nm and 609 nm, respectively.

**Fig. 2** exhibits the extinction spectra obtained for each preparation step, that is from the gold shell priming with 2 and 10 nm Au NPs to the complete shell growth. After the 2 nm and 10 nm Au NPs are grafted to the silica cores, a LSPR peak can be observed around 550 nm (dashed dark curve). Once a complete gold shell is grown, the LSPR peak is red-shifted to 624 nm for the 88 nm SiO$_2$ cores and to 977 nm for the 170 nm SiO$_2$ cores. Finally, once the growth process is completed, the Au shell leads to a blue-shift(Oldenburg et al., 1998, Loo et al., 2004) of the LPSR peak up to 584 nm and 609 nm for the 88 nm and 170 nm SiO$_2$ cores, respectively (see SEM images in **Fig. 1**). In order to assess the dimensions of the shell, Mie calculations were performed for SiO$_2$ cores of 90 nm and 170 nm diameters with varying gold shell thickness (see **Fig. 3**). These calculations indicate that a LSPR wavelength of 624 nm corresponds to a Au thickness of about 13 nm for 90 nm cores (which is in agreement with a 10 nm colloid layer around the SiO$_2$ cores). The 584 nm LSPR wavelength of the SiO$_2$@Au nanoparticles once the Au shell growth is completed corresponds to a shell thickness of about 20 nm which is also in agreement with the final dimensions of the SiO$_2$@Au nanoparticles as prepared. As for the larger SiO$_2$ cores, the two modes at 977 nm and 747 nm correspond to a shell thickness of 7 nm while the two modes at 795 nm and 609 nm correspond to a shell thickness of about 10 nm. It implies that for the larger nanoparticles, the growth was limited to the thickness of the 10 nm colloids. Moreover, the two extinction peaks when the process is completed (blue curve on **Fig. 2B**) are very broad indicating a large shell thickness distribution.

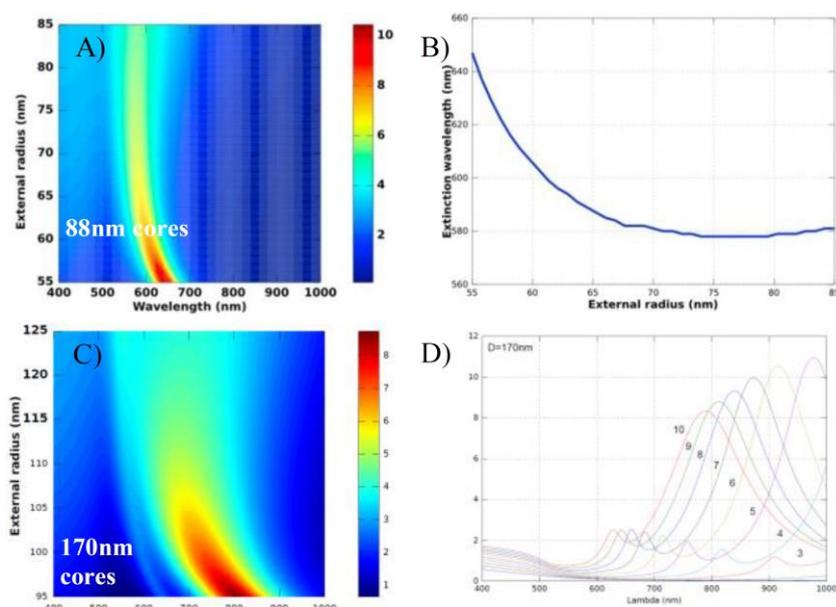

**Fig. 3** A) Visible-IR spectra of SiO$_2$@Au nanoparticles obtained from Mie calculations(Bohren and Huffman, 1983, Cai et al., 2010) for a core of diameter 90 nm with a gold shell of thickness 10 to 40 nm. A single LSPR peak can be observed. B) Extinction wavelength of the LSP Res-onance as a function of the external radius of the core-shell NP. The LSP peak blue-shifts when the Au shell thickness is increased. C) Visible-IR spectra of SiO$_2$@Au nanoparticles obtained from Mie calculations(Bohren and Huffman, 1983, Cai et al., 2010) for a core of diameter 170 nm with a gold shell of thickness 10 to 40 nm. Two LSPR peaks may be observed but the low wavelength one (around 600 nm) tends to decline with increasing shell thickness. D) Optical extinction spectra calculated for different shell thickness values (indicated in bold/nm) when the SiO$_2$ core diameter is 170 nm. The two LSP peaks can be clearly seen and blue-shift when the Au shell thickness is increased. The outside medium is supposed to be air.

In order to probe the potential of such core-shell NPs for sensing measurements, 50 μL droplets of 1,2-Bis(4-pyridyl)ethylene (BPE) molecules at varying concentrations were deposited on the surface of SiO$_2$@Au and 86 ± 6 nm Au NPs substrates12.The 88 nm core SiO$_2$@Au NPs were investigated because they exhibit a LSP resonance close to the 633 nm laser wavelength. The water droplet was then allowed to evaporate to ensure good adsorption of BPE molecules on the substrates. SERS measurements were then collected on LabRam instrument at 633 nm and 11 mW power. The data were acquired with a 10X objective. **Fig. 4** shows that the characteristic Raman peaks of BPE molecules can clearly be observed for $SiO_2$@Au NPs for concentrations down to $10^{-11}$ M. The comparison with the 86 nm Au NPs indicates that such core-shell structure helps in highly improving SERS sensitivity (by a factor of ten) with much lower gold quantity.

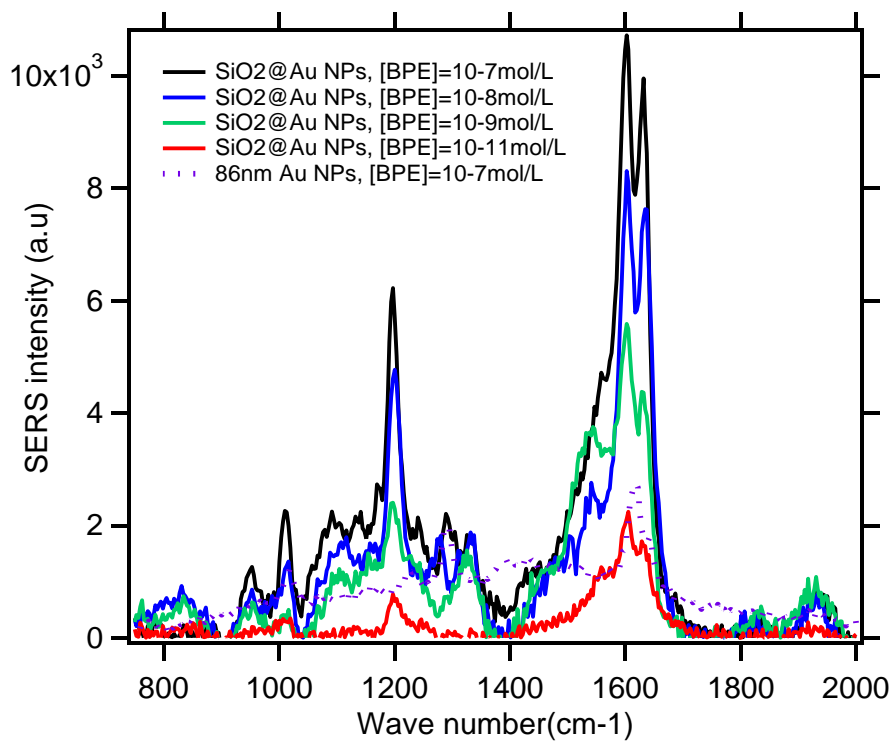

**Fig. 4** SERS spectra when BPE molecules of different concentrations are deposited onto glass substrates covered either with $SiO_2$@Au NPS or with 86 nm Au colloids.

To sum-up, in this paper, we detailed a chemical process for synthesizing $SiO_2$@Au nanoparticles with 88 nm and 170 nm $SiO_2$ cores and a complete Au shell. The process includes four steps: i) preparation of the silica cores, ii) grafting gold nano-particles over $SiO_2$ cores, iii) priming of the silica-coated gold nanoparticles with 2 and 10 nm gold colloids and finally iv) formation of a complete shell. SEM images combined to optical extinction measurements and calculations indicate that the Au shell growth was completed. This leads to two LSP modes for large $SiO_2$ NPs while the 88 nm ones exhibit only one extinction peak as expected from Mie calculations. The potential of such $SiO_2$@Au nanoparticles for biosensing was demonstrated via SERS measurements and comparison to 86 nm colloids. It is evidenced that such core-shell nanoparticles provides much more sensitive SERS detection of molecules with less gold matter than pure gold nanoparticles(Saini et al., 2014). Further directions of research should consist now in doping the $SiO_2$ cores with fluorescent molecules and studying the enhancement of the fluorescence signal induced by the Au shell.

**Acknowledgments**

Financial support of NanoMat (www.nanomat.eu) and NANOINKS by the "Ministère de l'enseignement supérieur et de la recherche," the "Conseil régional Champagne-Ardenne," the "Fonds Européen de Développement Régional (FEDER) fund," and the "Conseil général de l'Aube" is acknowledged. The authors thank the DRRT (Délégation Régionale à la Recherche et à la Technologie) of Champagne-Ardenne and the Labex ACTION project (contract ANR-11-LABX-01-01) for financial support. T. M. also thanks the CNRS via the chair « optical nanosensors ». I.I.L. thanks the "Conseil régional Champagne-Ardenne" for the postdoctoral fellowship.